\documentclass{article}

\usepackage{arxiv}

\usepackage[utf8]{inputenc} 
\usepackage[T1]{fontenc}    
\usepackage{hyperref}       
\usepackage{url}            
\usepackage{booktabs}       
\usepackage{amsfonts}       
\usepackage{nicefrac}       
\usepackage{microtype}      
\usepackage{lipsum}		
\usepackage{graphicx}
\usepackage{natbib}
\usepackage{doi}
\usepackage[linesnumbered,ruled,vlined]{algorithm2e}
\usepackage{multirow}

\title{Knowledge Transfer for Pseudo-code Generation from Low Resource Programming Language}

\author{ 
{Ankita  Sontakke}
        \\
	TCS Research\\
	\texttt{ankita.sontakke@tcs.com} \\
	\And
 {Kanika Kalra} \\
	TCS Research\\
	\texttt{kalra.kanika@tcs.com} \\
 \And
 {Manasi Patwardhan} \\
	TCS Research\\
	\texttt{manasi.patwardhan@tcs.com} \\
 \And
 {Lovekesh Vig} \\
	TCS Research\\
	\texttt{lovekesh.vig@tcs.com} \\
  \And
 {Raveendra Kumar Medicherla} \\
	TCS Research\\
	\texttt{raveendra.kumar@tcs.com} \\ 
  \And
  {Ravindra Naik} \\
	TCS Research\\
	\texttt{rd.naik@tcs.com} \\
  \And
  {Shrishti Pradhan} \\
	TCS Research\\
	\texttt{shrishti.pradhan@tcs.com} \\
}

\renewcommand{\shorttitle}{\textit{arXiv} Template}

\hypersetup{
pdftitle={A template for the arxiv style},
pdfsubject={q-bio.NC, q-bio.QM},
pdfauthor={David S.~Hippocampus, Elias D.~Striatum},
pdfkeywords={First keyword, Second keyword, More},
}

\begin{document}
\maketitle

\begin{abstract}
 Generation of pseudo-code descriptions of legacy source code for software maintenance is a manually intensive task. Recent encoder-decoder language models have shown promise for automating pseudo-code generation for high resource programming languages such as C++, but are heavily reliant on the availability of a large code-pseudocode corpus. Soliciting such pseudocode annotations for codes written in legacy programming languages (PL) is a time consuming and costly affair requiring a thorough understanding of the source PL. In this paper, we focus on transferring the knowledge acquired by the code-to-pseudocode neural model trained on a high resource PL (C++) using parallel code-pseudocode data. We aim to transfer this knowledge to a legacy PL (C) with no PL-pseudocode parallel data for training. To achieve this, we utilize an Iterative Back Translation (IBT) approach with a novel test-cases based filtration strategy, to adapt the trained C++-to-pseudocode model to C-to-pseudocode model. We observe an improvement of 23.27\% in the success rate of the generated C codes through back translation, over the successive IBT iteration, illustrating the efficacy of our approach.
\end{abstract}


\section{Introduction}
Low-level \emph{program design} expressed in natural language such as pseudocodes play a key role in several software engineering tasks such as system development, unit test case generation, and  system understanding. During greenfield application development, such low-level design is created by application designers. On the other hand, to rejuvenate or maintain an existing legacy system that has evolved over a period of time and deviated significantly from its original intended purpose, the system design should be recovered from the legacy system itself. However, design recovery \citep{biggerstaff1989design} is a difficult and time-consuming task due to inadequate functional knowledge of the application, outdated application documentation, and lack of skills in the legacy PL used. Inaccuracies in low level specifications due to manual errors while recovering them lead to incorrect new implementations, resulting in large financial risk to the organizations. This calls for a need of effectively recovering the low-level design in the form of \emph{pseudo codes} from legacy application programs automatically. Moreover, pseudocode being closer to the natural language, the task of automated pseudocode generation from codes would benefit other tasks such as code comment generation, serving as a pivot for code translation as well.  

Traditionally, the task of extracting low level design is performed \emph{manually} by developers with the help of code-assistants built using lightweight program analysis  \citep{niere2002towards,seemann1998pattern,biggerstaff1989design} or  statistical machine translation techniques \citep{Oda2015LearningTG,Rai2019GenerationOP, Fudaba2015PseudogenAT}. Recently, a few neural based techniques were proposed to build models that can map codes to their corresponding pseudocode \citep{Alhefdhi2018GeneratingPF, Yang2021FinegrainedPG, Yang2021DeepPseudoDP, Alokla2022RetrievalBasedTP}. To train such neural models is challenging as it requires a lot of parallel data in terms of code-pseudocode pairs. To cater to this need the community has come up with datasets such as SPoC \citep{Yang2021FinegrainedPG} and Django \citep{Oda2015LearningTG} for C++ and Python, respectively, on which the neural models are trained. However, for certain programming languages (PL) such as C or COBOL in which legacy codes are written, such datasets are unavailable. 
This highlights the need for  techniques which allow adaptation of these neural models to facilitate generation of pseudocodes for codes written in other legacy PLs with no availability of parallel training data.
\begin{figure} 
   \includegraphics[width=\textwidth]{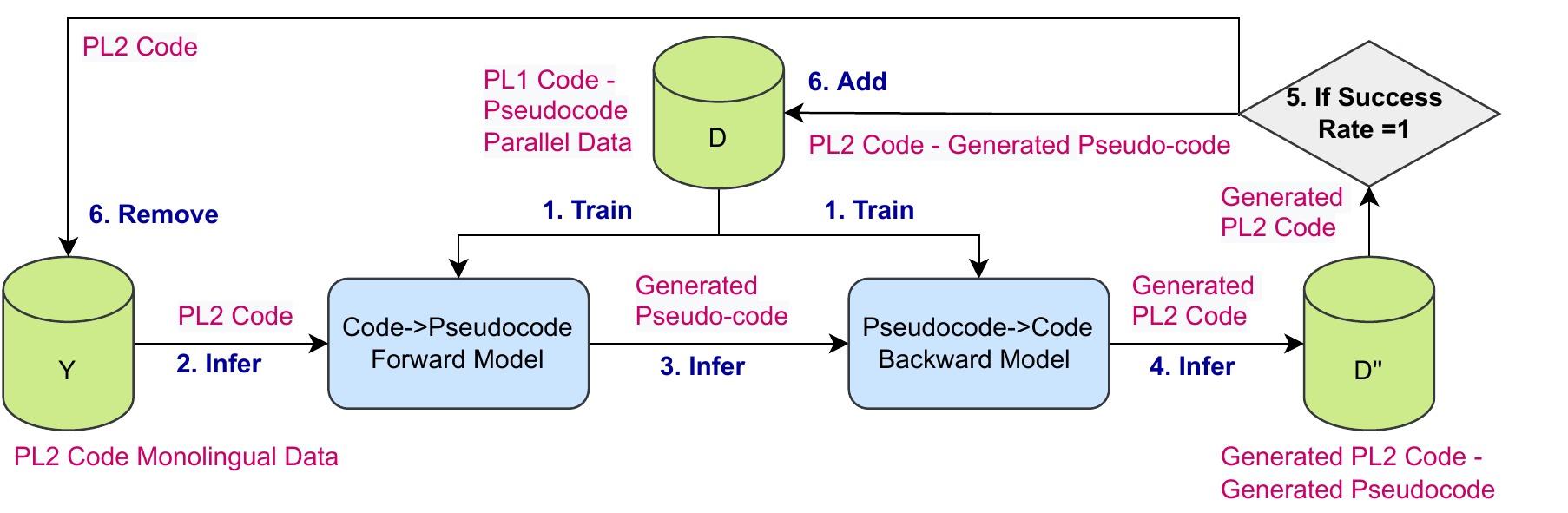}
   \caption{Iterative Back-Translation method for C programming language adaptation}
   \label{fig:backTrans}
 \end{figure}
 In this paper, we propose a promising direction to address this problem by using an Interactive Back Translation (IBT)  \citep{hoang-etal-2018-iterative} based approach (Figure \ref{fig:backTrans}). We apply IBT for the adaptation of the pseudocode generation model trained with codes written in C++ to generate pseudocode from C code snippets. The backward model is thus trained to replicate the original C program from the generated pseudo-code. The details are provided in Algorithm \ref{algo} and Section \ref{sec:app}. Improvements in the success rates of C codes generated by the backward (pseudocode-to-code) model from the pseudocode generated by the forward model (code-to-pseudocode), over the iterations of back translation demonstrates the efficacy of our approach. 
The main contributions of the paper are as follows:
\begin{itemize}
  \item State-of-the-art results for pseudocode generation from C++ codes with a model trained on SPoC parallel data \citep{kulal2019spoc}, with an improvement over \citep{Yang2021FinegrainedPG, Yang2021DeepPseudoDP, Alokla2022RetrievalBasedTP}. We call this a \textit{forward model} in our IBT approach.
  \item State-of-the-art results on the C++ code generation from  pseudocodes with a model trained on SPoC parallel data with an improvement over \citep{Xie2021ComposedFF, Shi2020IncrementalSW, Yasunaga2020GraphbasedSP, Kaan2021PseudocodeTC, Zhong2020SemanticSF}. We call this a \textit{backward model} in our IBT approach.
  \item 
  From the $0^{th}$ to the $1^{st}$ iteration of the IBT, we observe that additional 23.27\%  of C codes in the CodeNet dataset \citep{Puri2021ProjectCA}, generated by the \textit{backward model} from the  pseudocodes generated by the \textit{forward model}, execute correctly on all provided test-cases, thus  indicating superior pseudocodes generated for the source C codes. This showcases a promise in the IBT based approach for PL adaptation for the pseudocode generation task.

  \item To the best of our knowledge this is the first attempt at neural model adaptation for pseudocode generation.
\end{itemize}

\section{Related work}
\subsection{Code Summarization}
There are several neural based approaches which have addressed the task of summarizing code snippets using code-comment parallel data. 
Language Model (LM) based approaches such as PLBART \citep{ahmad2021unified}, CodeT5 \citep{wang2021codet5}, CoText \citep{phan2021cotext}, ProphetNet-Code \citep{qi2021prophetnet}, CodeTrans \citep{elnaggar2021codetrans}, and CodeBERT \citep{feng2020codebert}, UniXcoder \citep{Guo2022UniXcoderUC} pre-train a LM on mono-lingual PL data collected from Github and/or StackOverflow\footnote{https://stackoverflow.com/} with various pre-training objectives such as token masking, deletion, denoising, or infilling. They are further fine-tuned on code-summary pairs to learn code-text alignment and infer summaries for unseen codes. Approaches exploiting program analysis information use LSTMs \citep{hu2018deep, alon2018code2seq, leclair2019neural}, Transformers \citep{ahmad2020transformer,wu-etal-2021-code, zugner2021language, leclair2019neural, zhang2020retrieval}, Graph Neural Networks (GNNs) \citep{liu2020retrieval, leclair2020improved, wang2020learning} or a combination of these \citep{choi2021learning, shi2021cast} and inject program analysis information in the form of flattened Abstract Syntax Tress (ASTs) \citep{hu2018deep, alon2018code2seq, leclair2019neural} or Code Property Graphs (CPGs) \citep{liu2020retrieval} as inputs, or define attention cite{wu-etal-2021-code} \citep{liu2020retrieval} or relative positional encodings  \citep{zugner2021language}  between adjacent code tokens in ASTs and data and control flow graphs. 
The performance of these models in terms of BLEU scores on existing datasets of function-summary or code-comment pairs \citep{hu2018summarizing, wan2018improving, leclair2019neural, liu2020retrieval, husain2019codesearchnet, lu2021codexglue} is very low (in the range of 11.17 to 26.53) \citep{wang2021codet5}. This calls into question the utility of these models for real-life applications. Generation of pseudocode from source code, which is addressed in this paper, brings the code semantics closer to natural language and can be a step towards code summarization. More importantly, these approaches do not discuss the possibility of adapting a model trained on a given PL-summary pairs to a distinct PL, which is the main focus of this paper.   

\subsection{Code to Pseudocode Generation}
There are very few attempts to generate pseudocode from code. \citep{Yang2021FinegrainedPG, Alokla2022RetrievalBasedTP} generates pseudocode for C++ and Python programs using models trained on SPoC \citep{kulal2019spoc} and Django \citep{alhefdhi2018generating} parallel datasets. They use transformer and stack CNNs with gated linear unit (GLU) based code encodings to decode the pseudocode \citep{Yang2021FinegrainedPG} or use a retrieval based pseudocode generation approach  \citep{Alokla2022RetrievalBasedTP}. We surpass their results on the  C++-to-pseudocode generation task by using a CodeT5 \citep{wang2021codet5} sequence-to-sequence model with PL specific tokenization and pre-processing steps. Moreover, none of these approaches talk about adaptation of the models for an unseen PL, which is our main focus. 

\subsection{Pseudocode to Code Generation}
Auto-generation of code from pseudocode can facilitate a novice programmer to build on auto-generated codes given a low-level specification and thus can aid automated software development. \cite{dong2018coarse, zhong2017seq2sql} translate short text description into one-line program, whereas \cite{iyer2018mapping, rabinovich2017abstract} generate longer programs from NL descriptions and evaluate the generated code based on syntactic metrics like exact match or BLEU score. 
As opposed to this, LSTM \citep{kulal2019spoc, Yasunaga2020GraphbasedSP, Zhong2020SemanticSF} and Transformer \citep{Xie2021ComposedFF, Shi2020IncrementalSW} based line-by-line pseudocode-to-code generation approaches test the quality of the generated code with the available test-cases using the SPoC dataset \citep{kulal2019spoc}. 
\cite{kulal2019spoc} uses error localization methods to guide the search over the beam, whereas \cite{Zhong2020SemanticSF} further enhances search performance using semantic scaffolds and syntactic constraints with hierarchical beam search. DrRepair \citep{Yasunaga2020GraphbasedSP} repairs the generated code from pseudocode with a diagnostic feedback from compiler. 
\cite{Xie2021ComposedFF} propose the composed fine-tuning of pre-trained models to improve generalization for code generation. We not only surpass the results of these approaches for pseudocode-to-code generation, but also mainly use our model as the \textit{backward model} in IBT to facilitate evaluation and filtering of generated pseudocodes by the \textit{forward model} . 

\section{Problem Definition}
We have a parallel dataset $D = \{(c_v,w_v,p_v,t_v)\}^V_{v=1}$ where $c_v$ are C++ codes, $p_v$ are the pseudocodes, $w_v \in W$ are the worker ids annotating the pseudocodes and $t_v = \{(i_v,o_v)\}$ are input-output execution test cases. There is a one-to-one mapping between the code lines $\{c_v^l\}^{L_v}_{l=1}$ to the corresponding pseudocode lines $\{p_v^l\}^{L_v}_{l=1}$ $\forall (c_v, p_v)$, where $L_v$ are the total lines of code $c_v$. We have a monolingual data $Y = \{(c_u,t_u)\}^U_{u=1}$ where  $c_u$ are C codes and $t_u = \{(i_u,o_u)\}$ are the test cases for those codes. The task is to generate pseudocodes $p_u$ for the codes $c_u$ in $Y$, such that there is a one-to-one mapping between the code lines $\{c_u^l\}^{L_u}_{l=1}$ and the generated pseudocode lines $\{p_u^l\}^{L_u}_{l=1} \forall (c_u, p_u)$, where $L_u$ are the total lines of code $c_u$.

\section{Datasets}
As we are interested in generating pseudocodes for legacy C codes, we use C++-pseudocode parallel data provided by SPoC dataset \citep{kulal2019spoc} to train and test the forward code-to-pseudocode and backward pseudocode-to-code models. The intuition is that the overlap between the constructs of C and C++ PLs would facilitate the adaptation of the model trained on C++ for C codes. The SPoC dataset  provides C++ solutions for coding programs from codeforces, a competitive programming website \footnote{https://codeforces.com/}. Codeforces contains 18,356 C++ programs for 677 problem descriptions with the corresponding pseudocode and test cases. For each line of C++ code, the pseudocode is written by one or more of 59 crowd workers through Amazon Mechanical Turk platform\footnote{https://www.mturk.com/}.  
For evaluation, SPoC provides two test sets: (i) TESTP formed by splitting programs based on problem descriptions (158 problems with 1,820 programs) (ii) TESTW  formed by splitting programs based on workers (7 workers with 1,752 programs) to evaluate generalization. 
Remainder of the dataset is divided into train and validation sets in the ratio 90:10.

\begin{algorithm}
	\caption{IBT for PL Adaptation}
	\label{algo}
	\KwIn{Monolingual Code Data $Y = \{(c_u,t_u)\}^U_{u=1}$ where $c_u$ are C codes and $t_u = \{(i_u,o_u)\}$ are the test cases. $\{c_u^l\}^{L_u}_{l=1}$ are the code lines of $c_u$.
	\\\hspace{0.4in} Parallel data $D = \{(c_v,w_v,p_v,t_v)\}^V_{v=1}$ where  $c_v$ are C++ codes, $p_v$ are the 
	\\\hspace{0.4in} pseudocodes,  $w_v \in W$ are the worker ids writing the pseudocodes and $t_v = \{(i_v,o_v)\}$ 
	\\\hspace{0.4in} are the test cases. $\{c_v^l\}^{L_v}_{l=1}$ are the code lines of each code in $c_v$  with  the corresponding 
	\\\hspace{0.4in} pseudocode lines $\{p_v^l\}^{L_v}_{l=1}$. Thus, the samples of D are $\{(c_v^l,w_v,p_v^l)^{L_v}_{l=1}\}^V_{v=1}$ as all lines  
	\\\hspace{0.4in} of a code are annotated by the same worker.
	\\\hspace{0.4in} $B$: Beam Size, $B_d$: Budget, $I$: Number of Iterations
}	
	\KwOut{Trained forward C code-pseudocode $M_{c\rightarrow p}$ model}
	\For{$i\gets0$ \KwTo $I$}{
	Fine-tune forward code-to-pseudocode $M_{c\rightarrow p}$ model using $D$, with $w_u$ as prefixes per sample  \\
	Fine-tune backward pseudo-to-code $M_{c\leftarrow p}$ model using $D$ \\
    Feed $c_u^l \in c_u$ in $Y$ to $M_{c\rightarrow p}$ $\forall w_u \in W$ as prefixes to create $D' = \{(c_u^l, w_u,p_u^l,t_u)^{L_u}_{l=1}\}^{U*W}_{u=1}$ where $p_u^l \in p_u$ is the generated pseudocode line for code line $c_u^l \in c_u$
    \\
    Feed $p_u^l \in p_u$ in $D'$ to $M_{c\leftarrow p}$ to create $D'' = \{(p_u^l,w_u,{c'}_u^l,t_u)^{L_u}_{l=1}\}^{U*W}_{u=1}$ with beam size of $B$  for each line of generated codes ${c'}_u^l \in {c'}_u$ \\
    Use \textbf{Best-first-search} to search over $B$ beams of code lines $\{{c'}_u^l\}^{L_u}_{l=1} $ $ \forall {c'}_u \in D''$ with budget $B_d$ to get best prediction of code ${c'}_u$ \\ 
    \For{$c'_u \in D''$}{
    \If{ $\forall i_u \in t_u $ ${c'}_u(i_u) = o_u$ \tcp{Success Rate = 1, Filtration Strategy} }
    { $D = D \cup \{(c_u,w_u,p_u,t_u)\}$ \tcp{Data Augmentation} 
      $Y = Y - \{(c_u,t_u)\}$ \tcp{Test Set Update}
	}
    }
    \If{$Y$ = $\emptyset$}{$Break$}
    }
\end{algorithm} 

For PL adaptation, we use the C code dataset from Project CodeNet \citep{Puri2021ProjectCA}. We only take accepted (passed all the test cases) C codes into consideration. As SPoC dataset has programs solving simple problems  \citep{kulal2019spoc}, we choose 25,666 simple C CodeNet codes ($\approx$ 8.2\% of accepted C codes) along with their test cases as our mono-lingual PL data. The simple codes are the ones which have lower score given to the coding problem and more number of accepted submissions (ratio of accepted submissions (700) to the coding problem difficulty level (100) for AtCoder\footnote{https://atcoder.jp/} and codes for the coding problems having more than 2500 submission for AIZU\footnote{https://judge.u-aizu.ac.jp/onlinejudge/} from where CodeNet has collected the codes).

\begin{table}[t]
\caption{Success rates at budgets $B_d$ for pseudocode-to-code task on SPoC Test data
}
\label{table:pseudo-to-code result}
\centering
\begin{tabular}{l|lll|lll}
\hline
\textbf{Split} & \multicolumn{3}{c|}{\textbf{TESTP}} & \multicolumn{3}{c}{\textbf{TESTW}} \\ \hline
\textbf{Budget} & {$B_d$ = 1} & {$B_d$ = 10} & {$B_d$ = 100} & {$B_d$ = 1} & {$B_d$ = 10} & {$B_d$ = 100} \\ \hline
\cite{Shi2020IncrementalSW} & - & - & 32.5 & - & - & 51.0 \\
\cite{kulal2019spoc} & 17.8 & 28.4 & 34.2 & 30.7 & 44.4 & 53.7 \\
\cite{Yasunaga2020GraphbasedSP} & 17.8 & 31.4 & 38.5 & 30.7 & 48.0 & 57.0 \\
\cite{Xie2021ComposedFF} & 15.4 & - & - & 38.1 & - & - \\
\cite{Zhong2020SemanticSF} & 31.2 & 39.4 & 46.1 & 46.0 & 55.3 & 62.8 \\ \hline
\textbf{Ours} & \textbf{40.2} & \textbf{49.6} & \textbf{53.2} & \textbf{56.5} & \textbf{63.6} & \textbf{67.1} \\ \hline

\end{tabular}%
\end{table}
\begin{table}[t]
\caption{Scores for code-to-pseudocode task on SPoC Test data. *random splits.}
\label{table:Code-to-Pseudocode result}
\centering
\begin{tabular}{l|ll|cl}
\hline
Approach & 
\cite{Yang2021FinegrainedPG} & 
\cite{Alokla2022RetrievalBasedTP} & \multicolumn{1}{c}{Ours - TESTP} & \multicolumn{1}{c}{Ours - 
TESTW} \\ \hline
BLEU  & 46.45\textsuperscript{*}     & 50.28\textsuperscript{*}                     & 79.83                            & 64.06                            \\ \hline
\end{tabular}%
\end{table}

\section{Approach}\label{sec:app}
We use Iterative Back Translation (IBT) to transfer the knowledge of C++ code to pseudocode generation and repurpose it for C code to pseudocode generation. We accomplish this  by adapting the pseudocode generation model trained on the parallel data set $D$ of C++ codes using Iterative Back Translation and a test-cases based filtration strategy (Algorithm \ref{algo}).  We first fine-tune the \textit{forward} code-to-pseudocode ($M_{c\rightarrow p}$) model using $D$, with worker ids $w_u$ as prefixes to capture worker styles. We also fine-tune a \textit{backward} pseudocode-to-code ($M_{c\leftarrow p}$) model using the $D$ parallel data. We choose a Transformer based CodeT5 model  \citep{wang2021codet5} as our base model, which is pretrained on many natural language as well as PL specific tasks such as masked span prediction, masked identifier prediction, identifier tagging and bimodal dual generation, etc. We choose CodeT5 as it is trained using the CodeSearchNet \citep{husain2019codesearchnet} dataset which contains six PLs data including C and C\# and this prior knowledge of C PL would be useful for our downstream task. For both models, we use a C++ language specific tokenizer i.e clang python package \citep{kulal2019spoc} which makes the tokenization invariant to the set-of `space' characters used in the code lines. For example: the code lines, `\textit{else if (ans == int(ans))}' or `\textit{else if(  ans== int( ans))}' both tokenizes into consistent set of tokens`\textit{else if ( ans == int ( ans ) )}'. These tokenized code lines are further sub-tokenized using the CodeT5 tokenizer. We observe that, in the SPoC dataset, the C++ keyword `\textit{endl}' similar to "\textbackslash n" which inserts the newline character, is not annotated for the corresponding pseudocode by the crowd workers. As a pre-processing step, we modify the pseudocodes to include `\textit{print newline}' tokens for the corresponding `\textit{endl}' keyword in the C++ code. For example, the original pseudocode for `\textit{cout << - 1 << endl ;}' is `\textit{print -1}'. We modify it to `\textit{print - 1 print new line}'. With such pre-processed data we train the models in an auto-regressive fashion by using code and psuedo-code lines as parallel data, with cross entropy loss and a learning rate of $5e-6$. For the code-to-pseudocode model the epochs, input and output token lengths are 25, 128 and 200, respectively. For the pseudocode-to-code model the they are 50, 100 and 100, respectively, which covers most of the codes in the dataset. We choose the best models with least validation loss for inference . 

For the  \textit{forward model} we use beam size of 1 for the generated pseudocodes. For \textit{backward model}, we use a beam size $B$ of 10, that is for every generated pseudocode line from the \textit{forward model}, the \textit{backward model} generates 10 code lines. On the similar lines of  \cite{kulal2019spoc} we use \textbf{Best-first-search} to combine code lines forming the complete program and evaluate against the given $t_u$ test-cases. We first combine all the code lines on the top beams sequentially and execute the complete code with the test-cases. We replace the first error line from all compilation errors for a test-case with the next most probable code line present on the beam and re-execute the code. We assume $B_d$ as the budge in terms of number of allowable executions for a code.  If a program consisting of its code lines searched over the beam compiles successfully and passes all the test-cases within the budget then the success rate for that program is 1 otherwise it is 0.

For adaptation, we use monolingual C program dataset $Y$ and feed C code lines to the \textit{forward model} to generate pseudocode lines. The C programs are tokenized along the same lines as the C++ programs, as described earlier.  We append prefixes indicating worker ids $\in W$ to the code prior to generation. The top ten worker ids of workers who have annotated maximum number of code lines are utilized. We use the top-1 beam to obtain the the pseudocode from every worker. Thus, for every line of code we have 10 pseudocode lines generated. We feed these generated pseudocode lines to the \textit{backward model} to generate the code lines with a beam size of 10. To form a complete program from generated code lines, we use Best-first search algorithm  discussed above.
We choose budget $B_d$ as 10 (explained in Section \ref{sec:rad}). C codes executing correctly after backtranslation, along with the corresponding generated pseudocodes get added to the (parallel) training data $D$, and C codes that fail to execute remain in the monolingual C data $Y$ as test samples for the next iteration. In the subsequent iteration, we fine-tune both \textit{forward} and \textit{backward} models using the updated $D$ with a prefix indicating the PL (C++ or C) the program belongs to for better syntax discrimination and repeat the whole process. 
We execute the back translation pipeline for $I$ = 2 iterations to evaluate the efficacy of our proposed approach.


\begin{table}[t]
\caption{Ablation for the prepossessing steps NA: Not Applicable - Worker ID not used as prefix.}
\label{table:ablation}
\centering
\resizebox{\textwidth}{!}{%

\begin{tabular}{l|ll|llllll}
\hline
\multicolumn{1}{c|}{\multirow{3}{*}{Modification in Model}} & \multicolumn{2}{c|}{code-to-pseudocode} & \multicolumn{6}{c}{pseudocode-to-code} \\ \cline{2-9} 
\multicolumn{1}{c|}{} & \multicolumn{2}{c|}{BLEU} & \multicolumn{2}{c}{Exact Match} & \multicolumn{2}{c}{BLEU} & \multicolumn{2}{l}{Success Rate at B = 1} \\
\multicolumn{1}{c|}{} & TESTP & TESTW & TESTP & TESTW & TESTP & TESTW & TESTP & TESTW \\ \hline
CodeT5 tokenizer & 44.06 & 40.06 & 71.49 & 71.62 & 85.73 & 85.33 & 13.6 & 23.4 \\
+ PL specific tokenizer & 64.72 & 65.86 & 81.50 & 82.41 & 91.47 & 91.41 & 37.2 & 51.8 \\
+ Modified `\textit{endl}' keyword  & 65.33 & 65.97 & 82.66 & 83.72 & 92.04 & 92.02 & 40.2 & 56.5 \\
+Worker ID in input as prefix & 79.83 & 64.06 & NA & NA & NA & NA & NA & NA \\ \hline
\end{tabular}%
}
\end{table}

\begin{table}[t]
\caption{Iteration-wise Success rate of IBT for C adaptation}
\label{table:backTrans}
\centering
\resizebox{\textwidth}{!}{%
\begin{tabular}{lccc}
\hline
Iteration & \begin{tabular}[c]{@{}l@{}}\# Test Programs\end{tabular} & \begin{tabular}[c]{@{}l@{}}Success Rate at B=10\end{tabular} & \begin{tabular}[c]{@{}l@{}}Cumulative Success Rate\end{tabular} \\ \hline
Iteration 0 & 25666 & 60.84 \% & 60.84 \% \\
Iteration 1 & 10051 & 59.42 \% & \textbf{84.11 \%} \\ \hline
\end{tabular}%
}
\end{table}

\begin{table}[t]
\caption{Qualitative analysis of C-code Examples and generated Pseudo-codes over IBT iterations}
\label{table:qualitative analysis}
\resizebox{\textwidth}{!}{%
\begin{tabular}{lll}
\hline
1.& Original C code  & printf ( " \%d \%d \%d\textbackslash{}n " , a , c , b ) ;\\
&Iteration 0 Pseudo-code &  print " \% d \% d \% d \textbackslash n " , a , c and b\\
&Iteration 0 Top-1 C code & cout \textless{}\textless " \%d \%d \%d\textbackslash{}n " , a , c , b ;\\
&Iteration 1 Pseudo-code & print " \% d \% d \% d \textbackslash n " , a , c , b\\
&Iteration 1 Top-1 C code & printf ( " \%d \%d \%d\textbackslash{}n " , a , c , b ) ;\\
\hline
2.& Original C code  & putchar ( 10 ) ;\\
&Iteration 0 Pseudo-code & print 10\\
&Iteration 0 Top-1 C code & cout \textless{}\textless 10 \textless{}\textless ' \textbackslash{}n ' ;\\
&Iteration 1 Pseudo-code & print char 10\\
&Iteration 1 Top-1 C code & putchar ( 10 ) ; \\
\hline
3.& Original C code  & scanf ( " \%d " , \& x ) ;\\
&Iteration 0 Pseudo-code & \begin{tabular}[c]{@{}l@{}}call scanf with arguments \% d and address of x\end{tabular} \\
&Iteration 0 Top-1 C code & scanf ( \% d , \& x ) ;\\
&Iteration 1 Pseudo-code &\begin{tabular}[c]{@{}l@{}}call scanf with arguments \%d and \& x\end{tabular}\\
&Iteration 1 Top-1 C code &  scanf ( " \%d " , \& x ) ; \\
\hline
4.& Original C code  & fgets ( str , sizeof ( str ) , stdin ) ;\\
&Iteration 0 Pseudo-code & read str from the input to stdin\\
&Iteration 0 Top-1 C code & cin \textgreater{}\textgreater str ; \\
&Iteration 1 Pseudo-code & read str from the input to stdin\\
&Iteration 1 Top-1 C code & fgets ( str , sizeof ( str ) , stdin ) ; \\
\hline
5.& Original C code  & \begin{tabular}[c]{@{}l@{}} do \{ scanf ( " \%d " , \& s ) ; \}  while ( s \textless 0 || s \textgreater 100 ) ;\end{tabular}\\
&Iteration 0 Pseudo-code & \begin{tabular}[c]{@{}l@{}}while ( read \%d from stream , address of s is less than 0 or s is greater than 100 )\end{tabular} \\
&Iteration 0 Top-1 C code & while ( cin \textgreater{}\textgreater s , s \textless 0 || s \textgreater 100 )  \\
&Iteration 1 Pseudo-code & \begin{tabular}[c]{@{}l@{}}do \{ scanf ( " \% d " , address of s ) ; \} while s is less than 0 or s is greater than 100\end{tabular}\\
&Iteration 1 Top-1 C code &\begin{tabular}[c]{@{}l@{}}do \{ scanf ( " \%d " , \& s ) ; \}  while ( s \textless 0 || s \textgreater 100 ) ;\end{tabular} \\
\hline
6.& Original C code  &  scanf ( " \%d\%d\%d " , \& a , \& b , \& c ) ; \\
&Iteration 0 Pseudo-code & \begin{tabular}[c]{@{}l@{}}call scanf with arguments \%d \% d \% d  and \& a , \& b and \& c\end{tabular}\\
&Iteration 0 Top-1 C code & scanf ( \% d \% d \% d , \& a , \& b , \& c ) ; \\
&Iteration 1 Pseudo-code & scanf ( " \% d \% d \% d " , \& a , \& b , \& c )\\
&Iteration 1 Top-1 C code & scanf ( " \%d \%d \%d " , \& a , \& b , \& c ) ; \\
\hline

\end{tabular}%
}
\end{table}

\section{Results and Discussions}\label{sec:rad}
To evaluate our \textit{backward model} trained on SPoC pseudocode-to-C++ data, we use \textit{Success rate at B} metric \citep{kulal2019spoc}, which is the percentage of total programs that successfully compile and pass all available test-cases under the computation budget $B_d$. Computation budget $B_d$ is the number of times that a system can invoke the compiler and execute the compiled program on all test-cases. In short, we can search for maximum $B_d$ number of programs in the search space formulated by the beam. We evaluate our models by comparing our results with State-of-the-Art models for budget $B_d$ = 1, 10, 100 as shown in Table \ref{table:pseudo-to-code result}. We show that our model for C++ Code generation from pseudocode achieves success rates of 53.2\% and 67.1\% at $B$ = 100 on the SPoC TESTP and TESTW sets. 

Table \ref{table:Code-to-Pseudocode result} illustrate the comparison of our C++-to-Pseudocode (\textit{forward}) model on BLEU metric. We consider the SPoC original test splits TESTP and TESTW as opposed to the random splits performed by \cite{Yang2021DeepPseudoDP,Alokla2022RetrievalBasedTP}. Our model achieves state-of-the-art results on both TESTP and TESTW splits, proving the generalization capability of our approach as opposed to the benchmarks \citep{Yang2021DeepPseudoDP,Alokla2022RetrievalBasedTP}. 

We also conduct the ablation study for the distinct pre-processing steps applied on the code for both \textit{forward} and \textit{backward} models (Table \ref{table:ablation}). The PL specific tokenizer and modifying the pseudocode for the 'endl' keyword improves performance. Using crowd worker IDs as prefix for each code line helps the model understand the worker's annotation style and thus improves the BLEU score for TESTP which has overlapping worker ids with the train split. On the other hand, BLEU score decreases slightly for TESTW which has unseen workers. 

For the C PL codes we don't have the corresponding ground truth pseudo-code annotations available to test the performance of the \textit{forward model} for the pseudocode generation task, over the iterations of the back translation. We instead use the success rate for the C codes generated by the \textit{backward model} using the  pseudocodes generated by the \textit{forward model} from the original C codes, as part of the IBT, as the evaluation metric. Table \ref{table:backTrans} shows that the overall success rate of generated C codes increases by 23.27\% after only one iteration. This demonstrates the adaptation of the code-to-pseudocode model originally trained on C++ data to C codes over the IBT iterations. To further showcase the efficacy of our approach we illustrate the qualitative analysis of a few examples of C codes and generated pseudo-codes to demonstrate how the code-to-pseudocode model gets better at pseudocode generation over successive IBT iterations (Table \ref{table:qualitative analysis}). For example, in code line (1) the generated `\textit{Print}' pseudocode is mapped to C++ PL specific`\textit{cout}' syntax in iteration 0. However, it gets mapped to the C PL specific `\textit{printf}' function in iteration 1. This can be due to the C PL specific parallel data, which has been augmented in the training data, as a result of iteration 0, has the mapping of  `\textit{Print}' in pseudocode to `\textit{printf}' syntax, in some other composition. On similar lines, for codelines (2) and (3) the generated pseudocode gets better in iteration 1 by correctly reconstructing `\textit{putchar}' and `\textit{scanf}' constructs.  
For codeline (4) `\textit{fgets}' is mapped to read word correctly in iteration 0, even though the SPoC train set has not seen this construct, mainly de to the context. In iteration 1, the \textit{backward model} also learns to reconstruct the correct syntax from the pseudocode.
In codeline (5), the composition of \textit{scanf} and \textit{do while} construct unseen in the SPoC training data is learnt. Codeline (6) shows an example where the generated pseudocode is identical to the original code, in iteration 1, showcasing a failed pseudo-code generation case due to unseen \textit{scanf} construct. This qualitative analysis depicts how, in most of the cases, the \textit{forward model} gets better at the pseudocode generation task over the iterations of back translation, also leading to more correctly generated codes by the \textit{backward model}.

\section{Conclusion and Future Scope}
In this paper, we develop code-to-pseudocode and pseudocode-to-code models for C++ to provide State-of-the-Art results on SPoC dataset. We propose a promising new approach of adapting the C++ code-to-pseudocode model to C programs, with no parallel data, using IBT, where we utilize the available test cases to filter invalid code-pseudocode pairs. Our IBT pipeline demonstrates improvement in the execution rate of codes generated via back translation over successive iterations from the pseudocodes generated by the forward model. This showcases the efficacy of back-translation for adaptation of pseudocode generation to a legacy PL (C) with no parallel data when the original PL (C++) belongs to the same programming language family (some overlap in PL constructs) and has parallel data. 

For future work, we plan to use our IBT based approach for generated pseudocode adaptation to codes: (i) in the same PL with higher complexity  (ii) distinct legacy PLs (such as COBOL), having less overlap with the original PL (C++) in a few-shot (small parallel data) setting or by using a pre-defined common Intermediate Representation (IR), (iii) codifying implementations for distinct application domains (banking, retail, etc). We plan to define and use novel filtration strategies such as filtering (a) partial programs with exact match for certain code lines, or (b) codes with syntax errors checked using predefined restricted grammar, or (c) psuedocodes with no domain contents.  We also plan to initialize the models by meta-training with the available code-pseudocode parallel data belonging to multiple PLs such as C++  \citep{kulal2019spoc} and python \citep{alhefdhi2018generating},  to boost  performance for unseen PLs. All the above discussed scenarios are relevant in an industrial setting where it is infeasible to generate a lot of code-NL summary or pseudocode parallel data for distinct legacy programming languages and application domains.

\bibliographystyle{unsrtnat}
\bibliography{references}  

\begin{thebibliography}{46}
\providecommand{\natexlab}[1]{#1}
\providecommand{\url}[1]{\texttt{#1}}
\expandafter\ifx\csname urlstyle\endcsname\relax
  \providecommand{\doi}[1]{doi: #1}\else
  \providecommand{\doi}{doi: \begingroup \urlstyle{rm}\Url}\fi

\bibitem[Biggerstaff(1989)]{biggerstaff1989design}
Ted~J Biggerstaff.
\newblock Design recovery for maintenance and reuse.
\newblock \emph{Computer}, 22\penalty0 (7):\penalty0 36--49, 1989.

\bibitem[Niere et~al.(2002)Niere, Sch{\"a}fer, Wadsack, Wendehals, and
  Welsh]{niere2002towards}
J{\"o}rg Niere, Wilhelm Sch{\"a}fer, J{\"o}rg~P Wadsack, Lothar Wendehals, and
  Jim Welsh.
\newblock Towards pattern-based design recovery.
\newblock In \emph{Proceedings of the 24th international conference on Software
  engineering}, pages 338--348, 2002.

\bibitem[Seemann and von Gudenberg(1998)]{seemann1998pattern}
Jochen Seemann and J{\"u}rgen~Wolff von Gudenberg.
\newblock Pattern-based design recovery of java software.
\newblock \emph{ACM SIGSOFT Software Engineering Notes}, 23\penalty0
  (6):\penalty0 10--16, 1998.

\bibitem[Oda et~al.(2015)Oda, Fudaba, Neubig, Hata, Sakti, Toda, and
  Nakamura]{Oda2015LearningTG}
Yusuke Oda, Hiroyuki Fudaba, Graham Neubig, Hideaki Hata, Sakriani Sakti,
  Tomoki Toda, and Satoshi Nakamura.
\newblock Learning to generate pseudo-code from source code using statistical
  machine translation (t).
\newblock \emph{2015 30th IEEE/ACM International Conference on Automated
  Software Engineering (ASE)}, pages 574--584, 2015.

\bibitem[Rai and Gupta(2019)]{Rai2019GenerationOP}
Sawan Rai and Atul Gupta.
\newblock Generation of pseudo code from the python source code using
  rule-based machine translation.
\newblock \emph{ArXiv}, abs/1906.06117, 2019.

\bibitem[Fudaba et~al.(2015)Fudaba, Oda, Akabe, Neubig, Hata, Sakti, Toda, and
  Nakamura]{Fudaba2015PseudogenAT}
Hiroyuki Fudaba, Yusuke Oda, Koichi Akabe, Graham Neubig, Hideaki Hata,
  Sakriani Sakti, Tomoki Toda, and Satoshi Nakamura.
\newblock Pseudogen: A tool to automatically generate pseudo-code from source
  code.
\newblock \emph{2015 30th IEEE/ACM International Conference on Automated
  Software Engineering (ASE)}, pages 824--829, 2015.

\bibitem[Alhefdhi et~al.(2018{\natexlab{a}})Alhefdhi, Dam, Hata, and
  Ghose]{Alhefdhi2018GeneratingPF}
Abdulaziz Alhefdhi, Khanh~Hoa Dam, Hideaki Hata, and Aditya~K. Ghose.
\newblock Generating pseudo-code from source code using deep learning.
\newblock \emph{2018 25th Australasian Software Engineering Conference
  (ASWEC)}, pages 21--25, 2018{\natexlab{a}}.

\bibitem[Yang et~al.(2021{\natexlab{a}})Yang, Zhou, Chen, and
  Yu]{Yang2021FinegrainedPG}
Guang Yang, Yanlin Zhou, Xiang Chen, and Chi Yu.
\newblock Fine-grained pseudo-code generation method via code feature
  extraction and transformer.
\newblock \emph{2021 28th Asia-Pacific Software Engineering Conference
  (APSEC)}, pages 213--222, 2021{\natexlab{a}}.

\bibitem[Yang et~al.(2021{\natexlab{b}})Yang, Chen, Liu, and
  Yu]{Yang2021DeepPseudoDP}
Guang Yang, Xiang Chen, Ke~Liu, and Chi Yu.
\newblock Deeppseudo: Deep pseudo-code generation via transformer and code
  feature extraction.
\newblock \emph{ArXiv}, abs/2102.06360, 2021{\natexlab{b}}.

\bibitem[Alokla et~al.(2022)Alokla, Gad, Nazih, Aref, and
  Salem]{Alokla2022RetrievalBasedTP}
Anas Alokla, Walaa~K. Gad, Waleed Nazih, Mustafa Aref, and Abdel-Badeeh Salem.
\newblock Retrieval-based transformer pseudocode generation.
\newblock \emph{Mathematics}, 2022.

\bibitem[Hoang et~al.(2018)Hoang, Koehn, Haffari, and
  Cohn]{hoang-etal-2018-iterative}
Vu~Cong~Duy Hoang, Philipp Koehn, Gholamreza Haffari, and Trevor Cohn.
\newblock Iterative back-translation for neural machine translation.
\newblock In \emph{Proceedings of the 2nd Workshop on Neural Machine
  Translation and Generation}, pages 18--24, Melbourne, Australia, July 2018.
  Association for Computational Linguistics.
\newblock \doi{10.18653/v1/W18-2703}.
\newblock URL \url{https://aclanthology.org/W18-2703}.

\bibitem[Kulal et~al.(2019)Kulal, Pasupat, Chandra, Lee, Padon, Aiken, and
  Liang]{kulal2019spoc}
Sumith Kulal, Panupong Pasupat, Kartik Chandra, Mina Lee, Oded Padon, Alex
  Aiken, and Percy~S Liang.
\newblock Spoc: Search-based pseudocode to code.
\newblock \emph{Advances in Neural Information Processing Systems}, 32, 2019.

\bibitem[Xie et~al.(2021)Xie, Ma, and Liang]{Xie2021ComposedFF}
Sang~Michael Xie, Tengyu Ma, and Percy Liang.
\newblock Composed fine-tuning: Freezing pre-trained denoising autoencoders for
  improved generalization.
\newblock In \emph{ICML}, 2021.

\bibitem[Shi et~al.(2020)Shi, Bieber, and Sutton]{Shi2020IncrementalSW}
Kensen Shi, David Bieber, and Charles Sutton.
\newblock Incremental sampling without replacement for sequence models.
\newblock \emph{ArXiv}, abs/2002.09067, 2020.

\bibitem[Yasunaga and Liang(2020)]{Yasunaga2020GraphbasedSP}
Michihiro Yasunaga and Percy Liang.
\newblock Graph-based, self-supervised program repair from diagnostic feedback.
\newblock \emph{ArXiv}, abs/2005.10636, 2020.

\bibitem[Kaan et~al.(2021)Kaan, Ertas, Austin, and
  Brotman]{Kaan2021PseudocodeTC}
Kaan, Ertas, Austin, and Brotman.
\newblock Pseudocode to code translation using transformers.
\newblock In \emph{web.stanford.edu}, 2021.

\bibitem[Zhong et~al.(2020)Zhong, Stern, and Klein]{Zhong2020SemanticSF}
Ruiqi Zhong, Mitchell Stern, and Dan Klein.
\newblock Semantic scaffolds for pseudocode-to-code generation.
\newblock \emph{ArXiv}, abs/2005.05927, 2020.

\bibitem[Puri et~al.(2021)Puri, Kung, Janssen, Zhang, Domeniconi, Zolotov,
  Dolby, Chen, Choudhury, Decker, Thost, Buratti, Pujar, and
  Finkler]{Puri2021ProjectCA}
Ruchi Puri, David~S. Kung, Geert Janssen, Wei Zhang, Giacomo Domeniconi,
  Vladmir Zolotov, Julian Dolby, Jie Chen, Mihir~R. Choudhury, Lindsey Decker,
  Veronika Thost, Luca Buratti, Saurabh Pujar, and Ulrich Finkler.
\newblock Project codenet: A large-scale ai for code dataset for learning a
  diversity of coding tasks.
\newblock \emph{ArXiv}, abs/2105.12655, 2021.

\bibitem[Ahmad et~al.(2021)Ahmad, Chakraborty, Ray, and
  Chang]{ahmad2021unified}
Wasi~Uddin Ahmad, Saikat Chakraborty, Baishakhi Ray, and Kai-Wei Chang.
\newblock Unified pre-training for program understanding and generation.
\newblock \emph{arXiv preprint arXiv:2103.06333}, 2021.

\bibitem[Wang et~al.(2021)Wang, Wang, Joty, and Hoi]{wang2021codet5}
Yue Wang, Weishi Wang, Shafiq Joty, and Steven~CH Hoi.
\newblock Codet5: Identifier-aware unified pre-trained encoder-decoder models
  for code understanding and generation.
\newblock \emph{arXiv preprint arXiv:2109.00859}, 2021.

\bibitem[Phan et~al.(2021)Phan, Tran, Le, Nguyen, Anibal, Peltekian, and
  Ye]{phan2021cotext}
Long Phan, Hieu Tran, Daniel Le, Hieu Nguyen, James Anibal, Alec Peltekian, and
  Yanfang Ye.
\newblock Cotext: Multi-task learning with code-text transformer.
\newblock \emph{arXiv preprint arXiv:2105.08645}, 2021.

\bibitem[Qi et~al.(2021)Qi, Gong, Yan, Xu, Yao, Zhou, Cheng, Jiang, Chen,
  Zhang, et~al.]{qi2021prophetnet}
Weizhen Qi, Yeyun Gong, Yu~Yan, Can Xu, Bolun Yao, Bartuer Zhou, Biao Cheng,
  Daxin Jiang, Jiusheng Chen, Ruofei Zhang, et~al.
\newblock Prophetnet-x: Large-scale pre-training models for english, chinese,
  multi-lingual, dialog, and code generation.
\newblock \emph{arXiv preprint arXiv:2104.08006}, 2021.

\bibitem[Elnaggar et~al.(2021)Elnaggar, Ding, Jones, Gibbs, Feher, Angerer,
  Severini, Matthes, and Rost]{elnaggar2021codetrans}
Ahmed Elnaggar, Wei Ding, Llion Jones, Tom Gibbs, Tamas Feher, Christoph
  Angerer, Silvia Severini, Florian Matthes, and Burkhard Rost.
\newblock Codetrans: Towards cracking the language of silicon's code through
  self-supervised deep learning and high performance computing.
\newblock \emph{arXiv preprint arXiv:2104.02443}, 2021.

\bibitem[Feng et~al.(2020)Feng, Guo, Tang, Duan, Feng, Gong, Shou, Qin, Liu,
  Jiang, et~al.]{feng2020codebert}
Zhangyin Feng, Daya Guo, Duyu Tang, Nan Duan, Xiaocheng Feng, Ming Gong, Linjun
  Shou, Bing Qin, Ting Liu, Daxin Jiang, et~al.
\newblock Codebert: A pre-trained model for programming and natural languages.
\newblock \emph{arXiv preprint arXiv:2002.08155}, 2020.

\bibitem[Guo et~al.(2022)Guo, Lu, Duan, Wang, Zhou, and
  Yin]{Guo2022UniXcoderUC}
Daya Guo, Shuai Lu, Nan Duan, Yanlin Wang, Ming Zhou, and Jian Yin.
\newblock Unixcoder: Unified cross-modal pre-training for code representation.
\newblock In \emph{ACL}, 2022.

\bibitem[Hu et~al.(2018{\natexlab{a}})Hu, Li, Xia, Lo, and Jin]{hu2018deep}
Xing Hu, Ge~Li, Xin Xia, David Lo, and Zhi Jin.
\newblock Deep code comment generation.
\newblock In \emph{2018 IEEE/ACM 26th International Conference on Program
  Comprehension (ICPC)}, pages 200--20010. IEEE, 2018{\natexlab{a}}.

\bibitem[Alon et~al.(2018)Alon, Brody, Levy, and Yahav]{alon2018code2seq}
Uri Alon, Shaked Brody, Omer Levy, and Eran Yahav.
\newblock code2seq: Generating sequences from structured representations of
  code.
\newblock \emph{arXiv preprint arXiv:1808.01400}, 2018.

\bibitem[LeClair et~al.(2019)LeClair, Jiang, and McMillan]{leclair2019neural}
Alexander LeClair, Siyuan Jiang, and Collin McMillan.
\newblock A neural model for generating natural language summaries of program
  subroutines.
\newblock In \emph{2019 IEEE/ACM 41st International Conference on Software
  Engineering (ICSE)}, pages 795--806. IEEE, 2019.

\bibitem[Ahmad et~al.(2020)Ahmad, Chakraborty, Ray, and
  Chang]{ahmad2020transformer}
Wasi~Uddin Ahmad, Saikat Chakraborty, Baishakhi Ray, and Kai-Wei Chang.
\newblock A transformer-based approach for source code summarization.
\newblock \emph{arXiv preprint arXiv:2005.00653}, 2020.

\bibitem[Wu et~al.(2021)Wu, Zhao, and Zhang]{wu-etal-2021-code}
Hongqiu Wu, Hai Zhao, and Min Zhang.
\newblock Code summarization with structure-induced transformer.
\newblock In \emph{Findings of the Association for Computational Linguistics:
  ACL-IJCNLP 2021}, pages 1078--1090, Online, August 2021. Association for
  Computational Linguistics.
\newblock \doi{10.18653/v1/2021.findings-acl.93}.
\newblock URL \url{https://aclanthology.org/2021.findings-acl.93}.

\bibitem[Z{\"u}gner et~al.(2021)Z{\"u}gner, Kirschstein, Catasta, Leskovec, and
  G{\"u}nnemann]{zugner2021language}
Daniel Z{\"u}gner, Tobias Kirschstein, Michele Catasta, Jure Leskovec, and
  Stephan G{\"u}nnemann.
\newblock Language-agnostic representation learning of source code from
  structure and context.
\newblock \emph{arXiv preprint arXiv:2103.11318}, 2021.

\bibitem[Zhang et~al.(2020)Zhang, Wang, Zhang, Sun, and
  Liu]{zhang2020retrieval}
Jian Zhang, Xu~Wang, Hongyu Zhang, Hailong Sun, and Xudong Liu.
\newblock Retrieval-based neural source code summarization.
\newblock In \emph{2020 IEEE/ACM 42nd International Conference on Software
  Engineering (ICSE)}, pages 1385--1397. IEEE, 2020.

\bibitem[Liu et~al.(2020)Liu, Chen, Xie, Siow, and Liu]{liu2020retrieval}
Shangqing Liu, Yu~Chen, Xiaofei Xie, Jing~Kai Siow, and Yang Liu.
\newblock Retrieval-augmented generation for code summarization via hybrid gnn.
\newblock In \emph{International Conference on Learning Representations}, 2020.

\bibitem[LeClair et~al.(2020)LeClair, Haque, Wu, and
  McMillan]{leclair2020improved}
Alexander LeClair, Sakib Haque, Lingfei Wu, and Collin McMillan.
\newblock Improved code summarization via a graph neural network.
\newblock In \emph{Proceedings of the 28th International Conference on Program
  Comprehension}, pages 184--195, 2020.

\bibitem[Wang et~al.(2020)Wang, Zhang, Li, and Jin]{wang2020learning}
Wenhan Wang, Kechi Zhang, Ge~Li, and Zhi Jin.
\newblock Learning to represent programs with heterogeneous graphs.
\newblock \emph{arXiv preprint arXiv:2012.04188}, 2020.

\bibitem[Choi et~al.(2021)Choi, Bak, Na, and Lee]{choi2021learning}
YunSeok Choi, JinYeong Bak, CheolWon Na, and Jee-Hyong Lee.
\newblock Learning sequential and structural information for source code
  summarization.
\newblock In \emph{Findings of the Association for Computational Linguistics:
  ACL-IJCNLP 2021}, pages 2842--2851, 2021.

\bibitem[Shi et~al.(2021)Shi, Wang, Du, Zhang, Han, Zhang, and
  Sun]{shi2021cast}
Ensheng Shi, Yanlin Wang, Lun Du, Hongyu Zhang, Shi Han, Dongmei Zhang, and
  Hongbin Sun.
\newblock Cast: Enhancing code summarization with hierarchical splitting and
  reconstruction of abstract syntax trees.
\newblock \emph{arXiv preprint arXiv:2108.12987}, 2021.

\bibitem[Hu et~al.(2018{\natexlab{b}})Hu, Li, Xia, Lo, Lu, and
  Jin]{hu2018summarizing}
Xing Hu, Ge~Li, Xin Xia, D.~Lo, Shuai Lu, and Zhi Jin.
\newblock Summarizing source code with transferred api knowledge.
\newblock In \emph{International Joint Conference on Artificial Intelligence},
  2018{\natexlab{b}}.

\bibitem[Wan et~al.(2018)Wan, Zhao, Yang, Xu, Ying, Wu, and
  Yu]{wan2018improving}
Yao Wan, Zhou Zhao, Min Yang, Guandong Xu, Haochao Ying, Jian Wu, and Philip~S
  Yu.
\newblock Improving automatic source code summarization via deep reinforcement
  learning.
\newblock In \emph{Proceedings of the 33rd ACM/IEEE International Conference on
  Automated Software Engineering}, pages 397--407, 2018.

\bibitem[Husain et~al.(2019)Husain, Wu, Gazit, Allamanis, and
  Brockschmidt]{husain2019codesearchnet}
Hamel Husain, Ho-Hsiang Wu, Tiferet Gazit, Miltiadis Allamanis, and Marc
  Brockschmidt.
\newblock Codesearchnet challenge: Evaluating the state of semantic code
  search.
\newblock \emph{arXiv preprint arXiv:1909.09436}, 2019.

\bibitem[Lu et~al.(2021)Lu, Guo, Ren, Huang, Svyatkovskiy, Blanco, Clement,
  Drain, Jiang, Tang, et~al.]{lu2021codexglue}
Shuai Lu, Daya Guo, Shuo Ren, Junjie Huang, Alexey Svyatkovskiy, Ambrosio
  Blanco, Colin Clement, Dawn Drain, Daxin Jiang, Duyu Tang, et~al.
\newblock Codexglue: A machine learning benchmark dataset for code
  understanding and generation.
\newblock \emph{arXiv preprint arXiv:2102.04664}, 2021.

\bibitem[Alhefdhi et~al.(2018{\natexlab{b}})Alhefdhi, Dam, Hata, and
  Ghose]{alhefdhi2018generating}
Abdulaziz Alhefdhi, Hoa~Khanh Dam, Hideaki Hata, and Aditya Ghose.
\newblock Generating pseudo-code from source code using deep learning.
\newblock In \emph{2018 25th Australasian Software Engineering Conference
  (ASWEC)}, pages 21--25. IEEE, 2018{\natexlab{b}}.

\bibitem[Dong and Lapata(2018)]{dong2018coarse}
Li~Dong and Mirella Lapata.
\newblock Coarse-to-fine decoding for neural semantic parsing.
\newblock \emph{arXiv preprint arXiv:1805.04793}, 2018.

\bibitem[Zhong et~al.(2017)Zhong, Xiong, and Socher]{zhong2017seq2sql}
Victor Zhong, Caiming Xiong, and Richard Socher.
\newblock Seq2sql: Generating structured queries from natural language using
  reinforcement learning.
\newblock \emph{arXiv preprint arXiv:1709.00103}, 2017.

\bibitem[Iyer et~al.(2018)Iyer, Konstas, Cheung, and
  Zettlemoyer]{iyer2018mapping}
Srinivasan Iyer, Ioannis Konstas, Alvin Cheung, and Luke Zettlemoyer.
\newblock Mapping language to code in programmatic context.
\newblock \emph{arXiv preprint arXiv:1808.09588}, 2018.

\bibitem[Rabinovich et~al.(2017)Rabinovich, Stern, and
  Klein]{rabinovich2017abstract}
Maxim Rabinovich, Mitchell Stern, and Dan Klein.
\newblock Abstract syntax networks for code generation and semantic parsing.
\newblock \emph{arXiv preprint arXiv:1704.07535}, 2017.

\end{thebibliography}






\end{document}